\documentclass[sigconf,prl,superscriptaddress]{acmart}

\usepackage{xcolor}
\usepackage{mdframed}
\usepackage{multirow}
\usepackage[font=footnotesize]{caption}
\usepackage{enumitem}
\AtBeginDocument{%
  \providecommand\BibTeX{{%
    \normalfont B\kern-0.5em{\scshape i\kern-0.25em b}\kern-0.8em\TeX}}}

\newenvironment{gtheorem}%
  {\begin{mdframed}[backgroundcolor=lightgray]\begin{mdtheorem}{name}{label}}%
  {\end{mdtheorem}\end{mdframed}}

\copyrightyear{2021}
\acmYear{2021}
\setcopyright{acmcopyright}\acmConference[SPLC '21]{25th ACM International Systems and Software Product Line Conference - Volume A}{September 6--11, 2021}{Leicester, United Kingdom}
\acmBooktitle{25th ACM International Systems and Software Product Line Conference - Volume A (SPLC '21), September 6--11, 2021, Leicester, United Kingdom}
\acmPrice{15.00}
\acmDOI{10.1145/3461001.3473058}
\acmISBN{978-1-4503-8469-8/21/09}

\begin{document}

\title{Variability Fault Localization: A Benchmark}

\author{Kien-Tuan Ngo, Thu-Trang Nguyen, Son Nguyen, and Hieu Dinh Vo}

\email{{tuanngokien, trang.nguyen, sonnguyen, hieuvd}@vnu.edu.vn}
\affiliation{%
\institution{Faculty of Information Technology, VNU University of Engineering and Technology, Vietnam}
}
\begin{abstract}
Software fault localization is one of the most expensive, tedious, and time-consuming activities in program debugging. This activity becomes even much more challenging in Software Product Line (SPL) systems due to the variability of failures in SPL systems. These unexpected behaviors are caused by variability faults which can only be exposed under some combinations of system features. Although localizing bugs in non-configurable code has been investigated in-depth, variability fault localization in SPL systems still remains mostly unexplored. To approach this challenge, we propose a benchmark for variability fault localization with a large set of 1,570 buggy versions of six SPL systems and baseline variability fault localization performance results. Our hope is to engage the community to propose new and better approaches to the problem of variability fault localization in SPL systems.
\end{abstract}



\maketitle

\section{Introduction}
A Software Product Line (SPL) is the highly configurable system that enables developers to tailor the family of products from reusable software assets \cite{SPLBook}. 
This can be done by offering numerous \textit{features} controlled by \textit{options}. 
In Linux Kernel, there are +12K features, which is able to generate billions of scenarios.
Basically, a \textit{feature} is defined as a unit of functionality additional to the \textit{base software}. 
%
%For an SPL system, \textit{feature model (FM)} declares a set of features and impose the relationships between them (so-called constraints), where the presence or absence of some features requires or precludes the others.
%By configuring the statuses of enabling or disabling the features in the system, FM formalize different \textit{configurations} to satisfy desired feature combinations, which then applied to compose different program \textit{variants} (\textit{products}) \cite{?}. 
%
A set of \textit{selections} of all the features (\textit{configurations}) defines a \textit{product}.
The presence or absence of some features might require or preclude other features.
Feature dependencies are specified in a \textit{feature model} which constraints over features and defines valid configurations~\cite{SPLBook}.

In practice, to verify an SPL system, a subset of all of its valid products is selected for testing. In order to systematically test a system, various configuration sampling strategies have been proposed. Some popular sampling algorithms such as Combinatorial Interaction Testing \cite{kwise2012}, One-enabled~\cite{42bugs}, and One-disabled~\cite{42bugs} can be used for the configuration selection process. For each selected product, a test suite is generated to verify its behaviors.

However, the variability that is inherent to SPL systems challenges quality assurance (QA)~\cite{garvin2011feature, sampling_comparision, ase19prioritization}. 
In comparison with non-configurable code, finding bugs through testing in SPL systems is more problematic as a bug can be \textit{variable} (so-called \textit{variability bug}), which can only be exposed under some combinations of features~\cite{garvin2011feature, interaction_complexity}.
In other words, a system contains variability bugs if among the sampled products, some products pass all their tests while the others fail.
Hence, the buggy statements can only expose their bugginess in some particular products, yet cannot in others.

Despite the importance of finding variability bugs, the existing fault localization (FL) approaches are still limited. It is because these techniques are designed to find bugs in a particular product.
To isolate the bugs causing failures in multiple products of an SPL system, the slice-based methods~\cite{wong2016survey} could be used to identify the failure-related slices for each product independently of others. 
Consequently, there are a large number of statements in the whole system that need to be examined to find the bugs.  
This makes the slice-based methods become more impractical in SPL systems~\cite{wong2016survey}.
%

% Meanwhile, the state-of-the-art FL technique, Spectrum-Based Fault Localization (SBFL)~\cite{pearson2017evaluating, arrieta2018spectrum, spectrum_survey, keller2017critical, naish2011model}, assigns the suspiciousness scores to the statements based on the test execution information (i.e., program spectra) of each product independently of others. This also produces multiple ranked lists of the statements for a single system which failed by variability bugs. 
% %
% From these multiple ranked lists, developers cannot determine the right starting point to diagnose the root causes of the failures caused by the bugs. Hence, SBFL cannot be directly applied for variability bugs.

In addition, the state-of-the-art FL technique, Spectrum-Based Fault Localization (SBFL)~\cite{pearson2017evaluating, spectrum_survey, keller2017critical} cannot be directly applied for locating variability bugs. Indeed, SBFL assigns the suspiciousness scores to the statements based on the test execution information of each product independently of the others. For each product, it produces a ranked list of statements. As a result, there are multiple ranked lists for a single system which is failed by variability bugs. From these lists, developers cannot determine the right starting point to diagnose the root causes of the system failures.

%Adaptation
A naive solution to adapt SBFL for variability bugs in a system is that one can treat the whole system as a non-configurable code. This can be done by refactoring the mechanism controlling features in the system (e.g., \texttt{\#ifdef}) to the corresponding  \texttt{if-then} statements.
% and considering the states of the features as test inputs.
%
By this adaptation, for a faulty system, a single ranked list of the suspicious statements can be produced according to their suspiciousness scores. The score of each statement is measured based on the total number of the passing and failing tests executed by the statement in all the products.
However, this adaptation has two key problems which are caused by the incompatibility in testing of the different products.
First, \textit{in an SPL system, since the roles of a statement in different products are different, the statement behaves and is expected to behave differently in these products.}
% %
% %
Hence, the tests in a product, which are designed to verify the behaviors of the specific statements in this product, could not be used to verify the behaviors of those statements yet in another product.
% %
Consequently, counting all the tests in the different products to measure the suspiciousness of a statement could cause inaccurate assessments.
%
% \textit{First, counting all the tests in the different products to measure the suspiciousness of a statement could cause inaccurate assessments.} The reason is that in an SPL system, the roles of a statement in different products are different, the statement behaves and is expected to behave differently in these products. Therefore, the tests in a product, which are designed to verify the behaviors of the set of the specific statements in this product, could not be used to verify the behaviors of those statements yet in another product.
%
%
% Secondly, the score of a statement in the system tends to reflect more closely to the suspiciousness of that statement in the products which have more tests. This is because the total number of the tests in the system is counted from all the tested products. Therefore, if a product has more tests, it will have more impact on the suspiciousness scores of the statements in the whole buggy system.
% %
% This bias increases more clearly for the cases when the numbers of tests in the products are significantly different. 
% %
% As a consequence, the suspiciousness of a statement in the whole system might not be holistically measured. This could result in an inaccurate assessment in measuring statements' suspiciousness and lead to the inefficiency in localizing variability bugs.
%
Secondly, \textit{for a product having more tests, it will have more impact on the suspiciousness scores of the statements in the whole system.} This bias increases more clearly for the cases when the numbers of tests in the products are significantly different. It is because the total number of the tests used to measure the suspiciousness of statements in the system is counted from all the products. As a result, the suspiciousness of a statement in the whole system might not be holistically measured.
Thus, these two problems of this adaptation might cause the ineffectiveness of SBFL in localizing variability bugs.

To encourage the researchers and practitioners to propose better solutions for variability fault localization, we contribute a dataset including 1,570 buggy versions of 6 SPL systems with extensive test suites. In this dataset, there are  338 versions contain a single bug each and 1,232 versions contain multiple bugs.
% Based on the provided dataset, we expect the participants to propose new variability FL techniques. 
The proposed techniques should be evaluated using the following standard metrics: \textit{Rank}, \textit{EXAM} \cite{wong2008crosstab}, \textit{Recall at Top-$N$} \cite{lou2020can}, and \textit{PBL}~\cite{keller2017critical}, which are widely applied in the existing FL studies~\cite{keller2017critical, pearson2017evaluating, wong2016survey, spectrum_survey, wong2008crosstab}.

\vspace{2mm} 
\begin{gtheorem}
\textit{Variability Fault Localization Challenge}: 
Given a faulty SPL system containing variability bug(s) and a set of its sampled products with the test suites and test execution data, participants must propose new FL techniques to locate the buggy statement(s) in the system. The proposed techniques must be better than our baseline on the standard metrics in FL.
\end{gtheorem}
\vspace{2mm}

Our benchmark can be found at:

\vspace{2mm} 
\begin{gtheorem}
https://tuanngokien.github.io/splc2021/
\end{gtheorem}

\section{A Dataset of Variability Faults}
Since constructing a dataset of the real variability bugs with corresponding tests is a considerable task, until now there is no such public dataset.
The available datasets of variability bugs~\cite{42bugs,shiyi,98bugs} often lack of test information. 
Thus, they cannot be used to evaluate FL techniques which require test execution information.
Additionally, Just et al.~\cite{just2014mutants} has shown that the performance of FL techniques on real bugs can be estimated based on their performance on artificial bugs. 
That motivates us to propose a dataset of the artificial variability bugs with the corresponding tests which are systematically generated.
Furthermore, we categorize our dataset by different dimensions of bugs.
Based on that, the proposed FL techniques could be evaluated in some different circumstances.
%
% In particular, Section \ref{section:bug_generating_procedure} describes the process of producing test suites and applying mutation testing on software product lines for seeding variability-bugs. 
%
% Section \ref{section:subject_systems} presents the output buggy versions which are achieved from 6 different-sized SPL systems. 
%
% Finally, dataset structure and artifacts are reported in the Section \ref{section:dataset_artifacts}.

\subsection{Subject Systems}
\label{section:subject_systems}

%
% We aim to select the Java systems which have available features' source code and supported by composer \textit{FeatureHouse}~\cite{apel2011language} in a popular SPL system repository called \textit{SPL2Go}~\cite{?}.
% % \footnote{\textit{SPL2Go} - http://spl2go.cs.ovgu.de/}. 
% %
% %
% In \textit{SPL2Go}, some systems are not accepted because of existing issues in feature model or their implementation that make composer unable to build complete products' source code, or even raise some errors while compiling them.
% %
% Thus, we collected 6 Java SPL systems in \textit{SPL2Go} which are widely used in the existing studies in SPL systems and configurable code~\cite{apel2013strategies, al2016incling, interaction_complexity, apel2011language} to construct our dataset (Table \ref{table:subject_systems}).

We collected 6 Java SPL systems in \textit{SPL2Go}~\footnote{ http://spl2go.cs.ovgu.de/}, which are widely used in the existing studies about configurable code~\cite{apel2013strategies, interaction_complexity, apel2011language}, to construct our dataset (Table \ref{table:subject_systems}).
In addition, the products of each system are composed by \textit{FeatureHouse}~\cite{apel2011language}, a popular automated software composer. 
Indeed, there are some other systems in \textit{SPL2Go}, but they raise some errors while composing and compiling their products. Thus, they cannot be used in our dataset.
\subsection{Single Variability Bug Generation} 
\label{section:bug_generating_procedure}
%
% Previous studies \cite{?,?} have investigated the relationship between mutants and real faults in several programs, and found out there is no practically significant differences. 
%
% Some work also explores that the performance of FL techniques in real-world bugs can be estimated based on their performance on artificial bugs \cite{?,?}. 
%
% To construct the dataset of variability bugs, 
We design a process to systematically generate variability bugs, including three main steps as shown in Figure~\ref{fig:bug_generating_process_overview}: \textit{Product Sampling} and \textit{Test Generating}, \textit{Bug Seeding}, and \textit{Variability Bug Verifying}.

\textbf{Step 1 - Product Sampling and Test Generating}.
\label{section:sampling_and_test_generating}
Firstly, for an SPL system, a set of the products of the system is systematically sampled by the existing techniques~\cite{sampling_comparision, kwise2012}. Particularly, we sample a set of valid configurations based on the system's feature model with $4$-wise coverage by using \textit{SPLCA}~\cite{kwise2012}.
%
% With an SPL system, there are several ways to select variants which are suspiciously contain faulty interactions that causing failures during testing. 
%
% \textit{SPLCA}~\cite{kwise2012} employs greedy algorithm for generating covering array. 
% In this work, we use \textit{SPLCA}~\cite{kwise2012} to sample configurations with $4$-wise coverage. 
%
% Given a feature model, this tool is used to generate a set of valid configurations. 
%
For each configuration, a corresponding product is composed from the implementation of all the enabled features by using \textit{FeatureHouse} \cite{apel2011language}.
For each product $p$, a test suite is automatically generated using an existing test generation technique, \textit{Evosuite}~\cite{fraser2011evosuite}, to capture the original behaviors of $p$. 
%
% In this work, we use , which is a search-based test generator, to create tests.
For each test of $p$, the output will be recorded and used as the test oracle of the corresponding test in the product having the same configuration as $p$ in the mutated system in \textit{Step 2}.

%
% Since the lack of available test suites, we also need to create these test sets and use them to verify the given SPL system. 
%
% Currently, there is no testing techniques that is specialized for configurable system. 
%
% Thus, alternative strategy is to test all generated products individually. 
%
% In particular, Evosuite \cite{?}, a test case generator that achieved promising scores that mainly relying on search-based algorithm \cite{?,?}, is applied \cite{?} to automatically generate unit tests for each of products. 
%
% Additionally, with product P, each test output is recorded as the test oracle, and will be reuse to reveal failures in faulty version P' of P in the mutated system (\textit{Step 2}).

\textbf{Step 2 - Bug Seeding}.
To inject a fault into an SPL system, we randomly apply a modification to the original source code of the system by using a mutation operator. 
%
% Specially, multiple mutation operators, which are derived from mutation testing \cite{?}, are randomly seeded into system. 
%
In essence, the operator changes the original behaviors of several sampled products. The changes are expected to be captured by the products' generated tests. In other words, these products produce output that is different from the output recorded in \textit{Step 1}.
Particularly, we use $\mu$\textit{Java} tool~\cite{ma2006mujava} to create mutants at the statement-level of the system's source code.
We do not apply any operator which deletes a whole code statement.
% modify or delete a part of a statement. 
%
% The reason is that when a statement is removed from original code, this leads to complicating when evaluate FL techniques on ranked list.
Since when the whole statement is removed from the original code, the buggy statement which is expected to be localized by a FL technique might not be determined. Hence, we only use the operators modifying or deleting a part of a statement.

\textbf{Step 3 - Variability Bug Verifying}.
% \rule[-7pt]{0pt}{10pt}\\
%
In this step, we verify each generated bug to ensure that the fault is a variability bug and caught by the tests.
Particularly, for each mutated system, the collection of products, which are corresponding to the same set of the configurations sampled in \textit{Step 1}, are composed. 
After that, we run each of these products against its generated test suite.
A product is considered as a \textit{passing product} if it passes all the tests. 
In contrast, a product, which fails at least one test, is classified as a \textit{failing product}.
\textit{A bug is considered as a variability bug if it causes failures in certain sampled products}. In other words, after testing, among the sampled products, there are both passing and failing products.
Besides, during the testing process, test execution information will be recorded by a code coverage tool, \textit{OpenClover}~\footnote{https://openclover.org}.

\subsection{Multiple Variability Bugs Generation}
\begin{figure*}
    \centering
    \includegraphics[width=.63\linewidth]{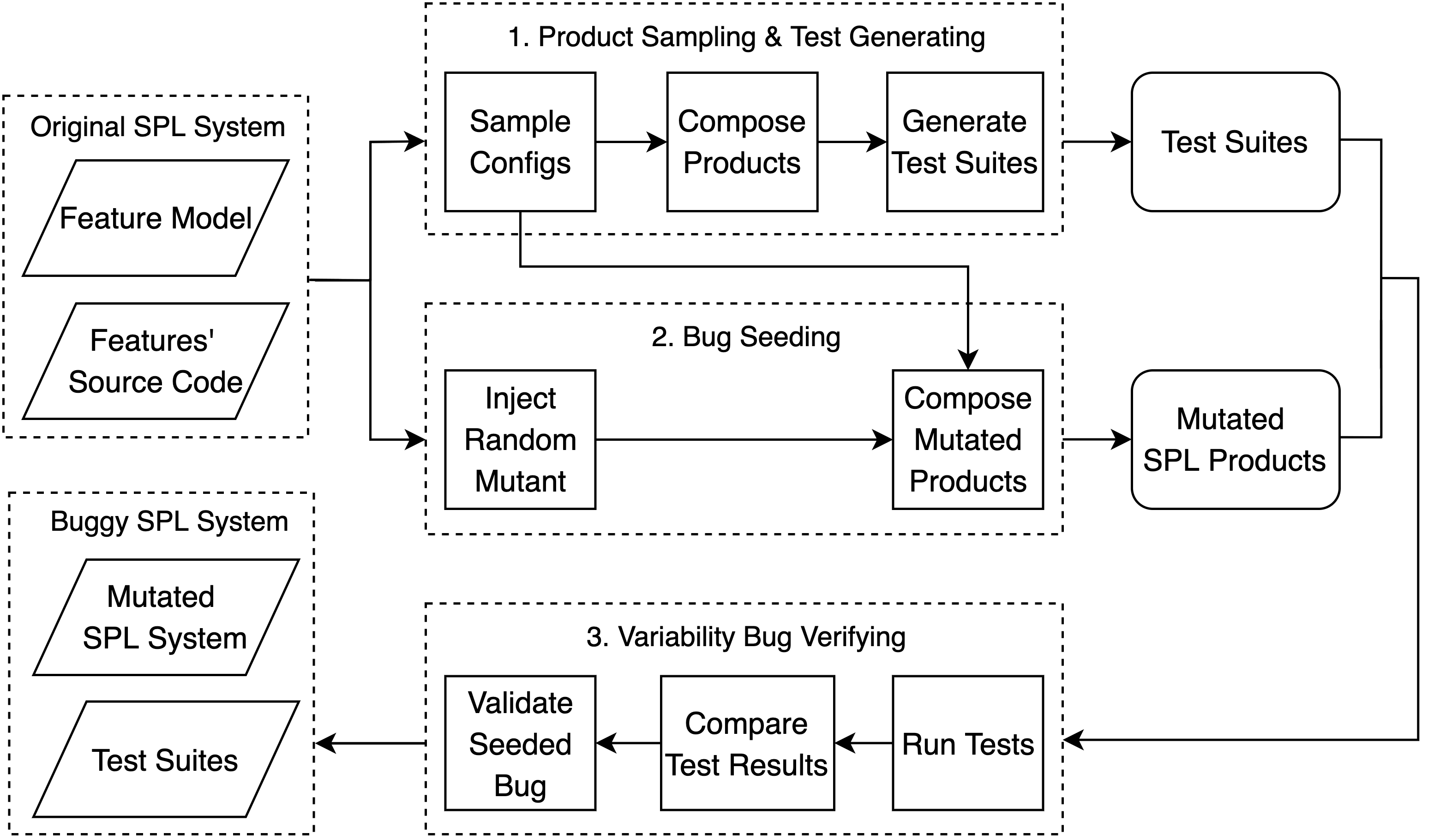}
    \caption{Bug-Generating Process Overview}
    \label{fig:bug_generating_process_overview}
\end{figure*}

For a more challenging setting, we create a dataset of the buggy systems which contain multiple variability bugs. Particularly, we extend \textit{Step 2} and \textit{Step 3} in Figure \ref{fig:bug_generating_process_overview} to generate such dataset: 

\textbf{Multiple Bugs Seeding}. In \textit{Step 2}, instead of applying a single modification, a random number, $n > 1$, of mutation operators are continuously used to mutate the source code of an SPL system.

\textbf{Variability Bugs Verifying}. In \textit{Step 3}, all the bugs need to be verified to ensure that they are variability bugs. Particularly, \textit{when one or more bugs are fixed, the remaining bugs still cause failures in certain products}. In other words, unless all the bugs are fixed, there still exists both the passing and failing products. We aim to simulate the bug-fixing process in practice. For each case, we gradually fix each bug by reverting the modification to the original state and do regression testing. 
The case would be accepted if there still exists both the passing and failing products until
all the bugs are fixed.

% Additionally, we aim to simulate the bug-fixing process in practice, that is, developer will repetitively apply FL technique to rank and repair the first examined buggy statement, then re-run regression testing until all the products pass all their tests.
% %
% So in case of multiple mutants are seeded into different locations, all the bugs need to be verified to guarantee variability bugs condition. 
% %
% Particularly, \textit{when one or more bugs are fixed, the remaining bugs still cause the failures in certain products}. 
% %
% To do that, for each case, we gradually fix each bug by reverting modified statement to its original state and then re-testing the whole system. 
% %
% The case would be accepted if there still exists both the passing and failing products until
% all the bugs are repaired.
%

\begin{table}
\caption{Dataset Statistics}
\label{table:subject_systems}
  \begin{tabular}{l|rr|rrr|rr}
  \toprule
    \multirow{2}{*}{\textbf{System}} & \multicolumn{2}{|c|}{\textbf{Details}} 
    & \multicolumn{3}{|c|}{\textbf{Test info}} & \multicolumn{2}{|c}{\textbf{Bug info}}\\ 
     & \#LOC & \#F & \#SP & \#Tests & Cov & \#V & \#IF \\
   \midrule
    ZipMe & 3460 & 13 & 25 & 255.0 & 42.9 & 304 & 2.7 \\
    GPL & 1944 & 27 & 99 & 86.9 & 99.4 & 372 & 13.0 \\
    Elevator-FH-JML & 854 & 6 & 18 & 166.0 & 92.9 & 122 & 3.6\\
    ExamDB & 513 & 8 & 8 & 133.3 & 99.5 & 263 & 1.1 \\
    Email-FH-JML & 439 & 9 & 27 & 86.0 & 97.7 & 126 & 4.1 \\
    BankAccountTP & 143 & 8 & 34 & 19.8 & 99.9 & 383 & 4.8\\
    \bottomrule
    % \multicolumn{8}{l}{\footnotesize \textbf{\#LOC} stands for number of lines of code.}\\
    % \multicolumn{8}{l}{\footnotesize \textbf{\#Features, \#SC} stand for number of features, and size of configurations sample, respectively.}\\
    \multicolumn{8}{l}{\footnotesize \textbf{\#F} and \textbf{\#SP} stand for the number of features and the average sample size.}\\
    % \multicolumn{8}{l}{\footnotesize \textbf{\#Test, \#Coverage} stand for average number of test cases per each product's test suite, and corresponding code coverage, respectively.}\\
    % \multicolumn{8}{l}{\footnotesize \textbf{\#Cases, \#IF} stand for number of buggy SPL system cases, and number of involving features relate to the bugs, respectively.}\\
    \multicolumn{8}{l}{\footnotesize \textbf{Cov} and \textbf{\#V} stand for the statement coverage (\%) and the number of buggy versions.}\\
    \multicolumn{8}{l}{\footnotesize \textbf{\#IF} stands for the average number of the involving features.}
  \end{tabular}
\end{table}

\subsection{Dataset of Variability Bugs}
\label{section:var_bug_dataset}

Table \ref{table:subject_systems} provides the general information of our dataset.
%About testing info
For each system, a set of products are sampled with 4-wise coverage. 
In general, for an SPL system, the number of the sampled products depends not only on the number of features but also the feature model of the system.
% For an SPL system, the number of the sampled products often depends on the number of features. Moreover, the sample size is also influenced by the feature model of the system.
% 
For instance, although \textsc{ExamDB} and \textsc{BankAccountTP} have the same number of features, to achieve 4-wise coverage by sampling technique, \textsc{BankAccountTP} needs to generate 34 products while this figure for \textsc{ExamDB} is only 8 products.

For generating tests, Table \ref{table:subject_systems} shows that the sizes of the generated test suites in different systems are different. For example, more than 8.5K test cases are created for \textsc{GPL} in total.
In our dataset, there are 5/6 systems whose generated test suite reaches +90\% statement coverage. Moreover, three of them almost reach 100\% statement coverage.
Especially due to a large code base, \textsc{ZipMe} has 255 tests per product, but its statement coverage only stays at 42.9\%. 
%

%GENERAL BUG INFO
As seen in Table \ref{table:cases_categorized_by_number_of_bugs}, we generated 1,570 buggy versions of the subject systems. Among them, 338 versions contain a single bug each, while 1,232 versions have two or more bugs.
In the dataset, the number of bugs might not be proportional to the size of systems.
For instance, \textsc{ZipMe} contains a larger number of statements and has more features than \textsc{ExamDB}. However, there are more bugs generated in \textsc{ExamDB}. 
The reason is, the number of mutation operators applicable to  \textsc{ExamDB} is greater. Thus, there are more mutants and variability bugs that can be generated in \textsc{ExamDB} than in \textsc{ZipMe}.
Moreover, the quality of test suite also plays a critical role in generating variability bugs. The test suite with higher statement coverage is more effective in detecting the unexpected behaviors caused by the seeded bugs in the system. 
Hence, for a system with a better test suite, there might be more variability bugs accepted.
%
% For example, in single-bug context, \textsc{ZipMe} actually able to create up to 450 mutated versions, but it remains only 56 cases that contain variability-bugs due to the low of test coverage. 

%
% Table \cite{?} also represents that number of cases have multiple bugs often bigger than the single-bug one. The reason is ...

%BUGS BY MUTATION OPERATORS
%
\begin{figure}
    \centering
        \includegraphics[width=.55\linewidth]{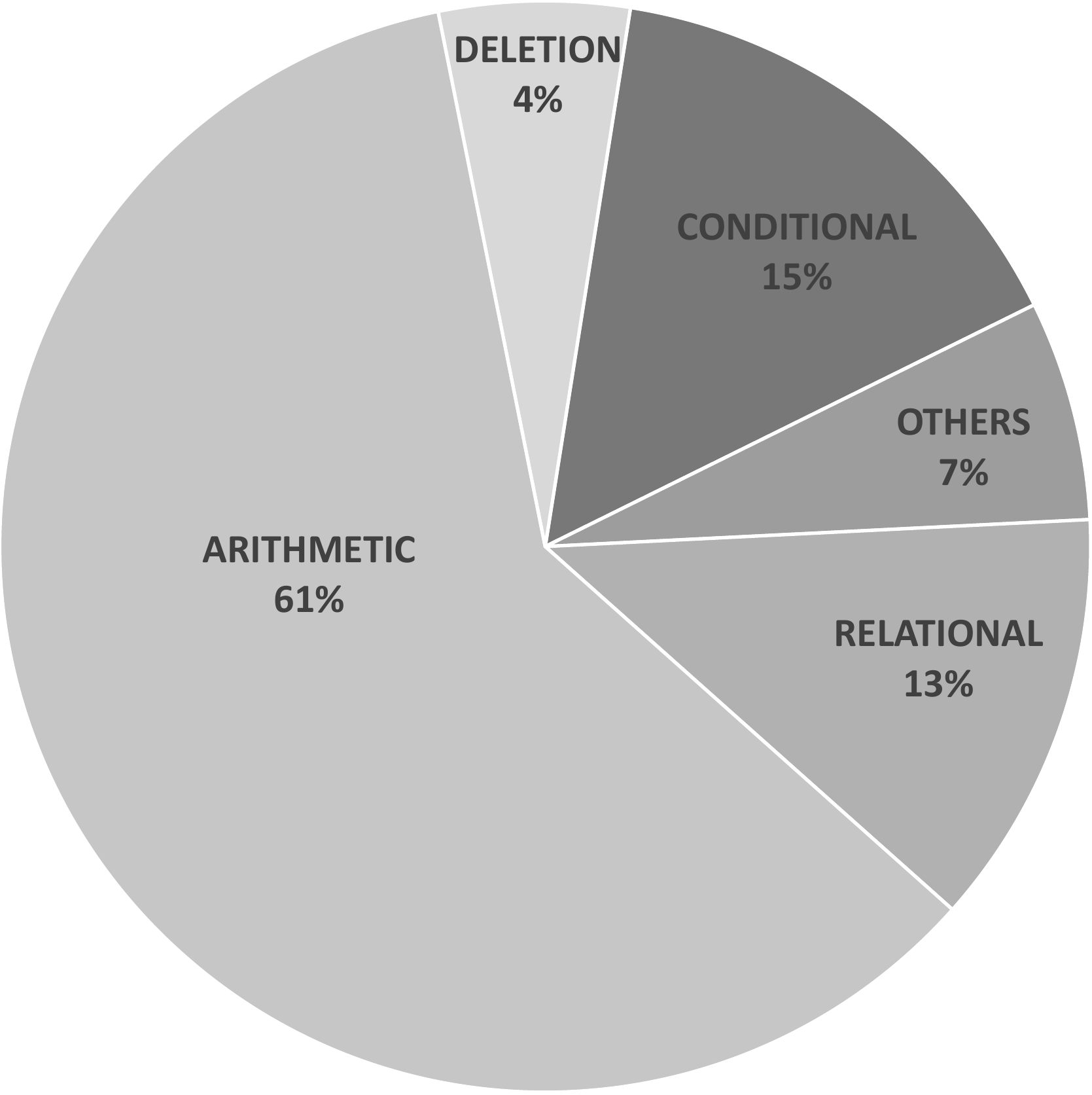}
    \caption{Variability Bugs by Applied Mutation Operators}
    \label{fig:bugs_by_mutation_operators}
\end{figure}

\textbf{Variability Bugs by Applied Mutation Operators}.
Figure \ref{fig:bugs_by_mutation_operators} shows the proportions of the bugs categorized by the groups of mutation operators~\cite{ma2005description}. 
As seen, there are more than haft of them generated by using \textit{Arithmetic} group.
% , with \textit{Conditional} (548) and \textit{Relational} (352) coming in the second and third, respectively.  
%
The reason is, comparing to the other groups, \textit{Arithmetic} group contains more mutation operators, such as \textit{AODS}, \textit{AODU}, \textit{AOIS}, \textit{AOIU}, and \textit{AORS}~\cite{ma2005description}, that are applicable for mutating the source code of our selected  systems.
%
% In contrast, \textit{Deletion} contains least operators than \textit{Arithmetic}. 
%
% Meanwhile, the number of the bugs by applying \textit{Deletion} group is less than the numbers of bugs generated by others. 
% This is because these bugs are more challenging to be detected than one generated other groups~\cite{?}.
%
% Thus, the number of variability bugs generated by using \textit{Deletion} operators is fewer.
%

\begin{figure}
    \centering
    \includegraphics[width=.55\linewidth]{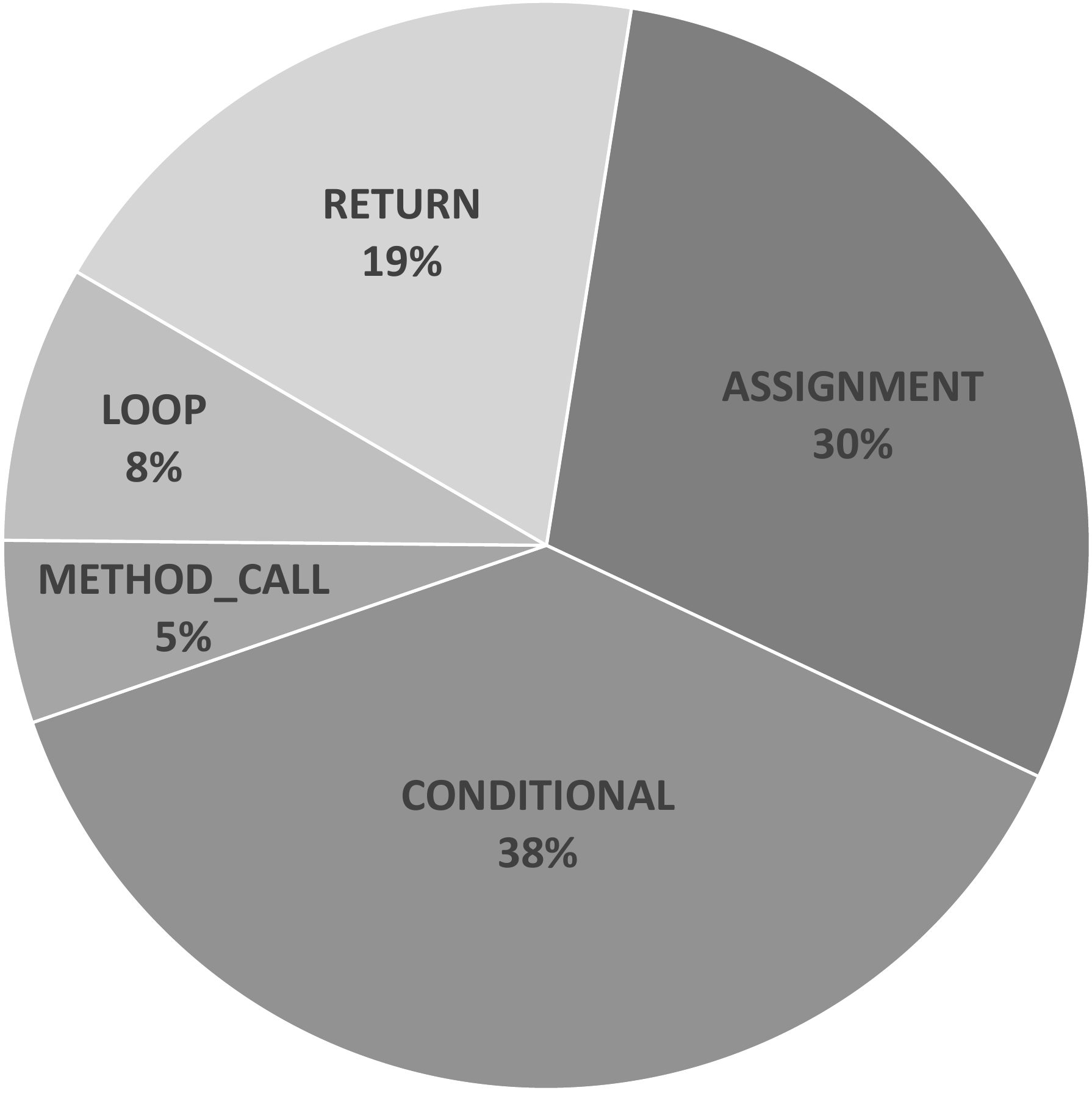}
    \caption{Variability Bugs by Code Elements}
    \label{fig:bugs_by_code_elements}
\end{figure}

%BUGS BY CODE ELEMENTS
\textbf{Variability Bugs by Code Elements}.
Figure \ref{fig:bugs_by_code_elements} shows the proportions of the bugs classified by code element types~\cite{sobreira2018dissection}.
% including \textit{Assignment}, \textit{Conditional}, \textit{Loop}, \textit{Method Call} and \textit{Return}. 
%
As seen, groups of \textit{Conditional} and \textit{Assignment} contain more bugs than the others. These proportions are similar to the distribution reported by the prior study on the popular real-fault repository~\cite{sobreira2018dissection}.

%BUGS BY INVOLVING FEATURES
\textbf{Variability Bugs by Involving Features}.
Furthermore, for a specific SPL system, the performance of FL techniques might be influenced by the number of features which must be \textit{actually} enabled/disabled to reveal the bugs (\textit{involving features}).
In this work, a feature is an involving feature to a variability bug if from a failing product, when switching its current selection (the state of being on or off) makes the resulting product pass all its tests. 
If the resulting product has not been sampled, we additionally compose the product and generate its tests.
%
% Note that this strategy only works in single-bug cases.
%
For a system containing multiple bugs, since the failures in the sampled products are caused by various bugs, detecting the involving features for each bug in this case might be impossible.
For a particular bug, when one of its involving features is switched, the resulting product might still fail because of the other bugs.
Thus, in our dataset, we only categorized the buggy versions containing a single bug by the number of involving features.
% For multiple bugs, the fact is that a test suite need to capture unexpected behaviors produced by all of them.
%
% Thus, switching an \textit{actually} involving feature of one bug might still make resulting product fails at the tests that verify other bugs' behaviors, and then corresponds to a failing product.
%
% For accurately finding the set of involving features, we only apply the process to single-bug versions.
%
In our dataset, the number of involving features is in the range of [1, 25], about 76\% of the cases are less than or equal to 7.

\begin{table}
\caption{Buggy versions categorized by number of contained bugs}
\label{table:cases_categorized_by_number_of_bugs}
  \begin{tabular}{l|r|r|r}
  \toprule
     \textbf{System} & \textbf{Single-Bug} & \textbf{2-Bug} & \textbf{3-Bug}\\
   \midrule
    ZipMe & 55 & 120 & 129\\
    GPL & 105 & 190 & 77\\
    Elevator-FH-JML & 20 & 41 & 61\\
    ExamDB & 49 & 126 & 88\\
    Email-FH-JML & 36 & 34 & 56\\
    BankAccountTP & 73 & 238 & 72\\
    \bottomrule
  \end{tabular}
\end{table}
\subsection{Description of Dataset Artifacts}
\label{section:dataset_artifacts}

Our dataset is published on our website with all required information to evaluate participants' solutions. In particular, all the cases are organized in different folders and each represents a version of the entire SPL system. In each case, artifact structure is as below:

\begin{itemize}[leftmargin=*]
\item \textbf{Feature Model:} 
% Given SPL system always contains the feature model and used in the sampling process. 
The feature model is in GUIDSL format~\cite{batory2005feature}.

\item \textbf{Sampled Configurations:} A set of configurations with $4$-wise coverage which are saved in different files.

\item \textbf{Source Code:} The system and all the composed products. Note that each product is composed based on feature superimposition mechanism of \textit{FeatureHouse} ~\cite{apel2011language}. For remapping purposes, the link between each code statement in a product and the corresponding statement in the system is recorded.

\item \textbf{Test Cases:} All the generated test cases of each sampled product.

\item \textbf{Bug Report:} The locations of modified statements in the system.

\item \textbf{Test Execution:} This data is supplied in the execution log, informs how many times each statement is executed for each test. 
% In general, FL techniques can use this data as an input to localize the bugs.   
\end{itemize}

\section{Solution Evaluation}

This section describes several standard metrics to evaluate FL techniques for variability bugs in SPL systems. Furthermore, we present the results of the naive adaptation of SBFL technique on the proposed dataset as the baseline results of this challenge.

\subsection{Evaluation Metrics}
The main focus of FL is to help developers find a good starting point to inspect and initiate the bug-fixing process. Therefore, the effectiveness of FL technique generally based on the percentage of code that needs to be examined until the first faulty location is found. In this challenge, we apply the standard metrics which are widely used in evaluating FL techniques~\cite{keller2017critical, pearson2017evaluating, wong2016survey, spectrum_survey, wong2008crosstab,lou2020can}.

\textbf{Rank.} 
% \textit{Rank} is the position of a buggy statement in the ranked list. 
The lower \textit{Rank}, the better approach. If there are multiple statements having the same scores, buggy statements are ranked last among them.
For the cases of multiple bugs, we measured \textit{Rank} by the position of the first bug in the ranked lists.

\textbf{EXAM.} \textit{EXAM} \cite{wong2008crosstab} is the percentage of the statements that must be examined until the first faulty statement is reached:\\
\begin{displaymath}
  EXAM = \frac{\text{Number of examined statements}}{\text{Total number of statements}} \times 100\%
\end{displaymath}\vspace*{0mm}

\textbf{Recall at Top-$N$ (Top-$N$)}. \textit{Top}-$N$~\cite{lou2020can} was devised to report the number of cases that at least one bug was found after examining $N$ statements in the ranked list. 
% In practice, developers only investigate the top-ranked suspicious statements before giving up. In this benchmark, we consider $N \in [1, 5]$.

\textbf{Proportion of Bugs Localized (PBL).} \textit{PBL}~\cite{keller2017critical} measures the proportion of the bugs which are detected after examining a certain number of the statements in the system. 

% Please note that, when multiple statements have equal suspiciousness score and correspond to the same position in the ranked list, FL technique must assess the buggy statements as the worst rank between them.

\subsection{Baseline Results}

% To adapt the state-of-the-art FL technique, SBFL, for localizing variability bugs in an SPL system, the whole system can be treated as a non-configurable code. 
To construct baseline results, we conducted several experiments with the naive adaption of SBFL which considers the whole SPL system as a non-configurable code. The suspiciousness score of a statement is measured based on the tests counted from all the products. We use this adaption with different SBFL metrics~\cite{wong2016survey,spectrum_survey} to evaluate the baseline performance on localizing variability bugs in both single-bug and multiple-bug settings.

\subsubsection{Performance in localizing single bug}

\begin{table*}
\caption{SBFL Performance in Single-bug Setting}
\label{table:baseline_SBFL_single_bug}
  \begin{tabular}{l|rr|rr|rr|rr|rr|rr}
  \toprule
\multirow{2}{*}{\textbf{Metric}} & \multicolumn{2}{c|}{\textsc{\textbf{ZipMe}}} & \multicolumn{2}{c|}{\textsc{\textbf{GPL}}} & \multicolumn{2}{c|}{\textsc{\textbf{Elevator}}} & \multicolumn{2}{c|}{\textsc{\textbf{ExamDB}}} & \multicolumn{2}{c|}{\textsc{\textbf{Email}}} & \multicolumn{2}{c}{\textsc{\textbf{BankAccountTP}}} \\
 & Rank & EXAM & Rank & EXAM & Rank & EXAM & Rank & EXAM & Rank & EXAM & Rank & EXAM \\
 \midrule
    Taratula & 23.98 & 1.03 & 10.36 & 1.07 & 18.40 & 4.11 & 5.61 & 2.24 & 13.81 & 5.59 & 5.53 & 7.23 \\
    Ochiai & 18.40 & 0.79 & 9.09 & 0.94 & 8.55 & 1.91 & 3.31 & 1.32 & 4.56 & 1.84 & 3.95 & 5.15 \\
    \textbf{Op2} & \textbf{12.67} & \textbf{0.55} & \textbf{8.86} & \textbf{0.92} & \textbf{4.25} & \textbf{0.95} & \textbf{3.24} & \textbf{1.29} & \textbf{4.03} & \textbf{1.63} & \textbf{3.58} & \textbf{4.66} \\
    Barinel & 23.98 & 1.03 & 10.36 & 1.07 & 18.40 & 4.11 & 5.61 & 2.24 & 13.81 & 5.59 & 5.55 & 7.24 \\
    Dstar & 18.20 & 0.78 & 9.09 & 0.94 & 8.40 & 1.88 & 3.29 & 1.31 & 4.61 & 1.87 & 3.92 & 5.11 \\
  \bottomrule
    \multicolumn{13}{l}{\footnotesize \textsc{\textbf{Elevator}}, \textsc{\textbf{Email}} are the abbreviations for \textsc{\textbf{Elevator-FH-JML}} and \textsc{\textbf{Email-FH-JML}} systems, respectively.}
  \end{tabular}
\end{table*}

In this experiment, we use different SBFL metrics to localize the buggy statements in 338 cases where each case contains only one bug. Table~\ref{table:baseline_SBFL_single_bug} shows
the average \textit{Rank} and \textit{EXAM} of SBFL using the 5 most popular SBFL metrics. The results of other SBFL metrics can be found on our website~\cite{website}.
As seen, there are several SBFL metrics that obtained quite similar performance, such as Tarantula and Barinel, Ochiai and Dstar, etc. 
Op2 is the most effective metric which achieves significantly better performance in both \textit{Rank} and \textit{EXAM} compared to the others' results. Interestingly, the average \textit{Rank} with Op2 in most of the systems is around $6^{th}$. Meanwhile, the average \textit{Rank} in \textsc{ZipMe} is about $13^{th}$, perhaps because of the low-quality test suites.
%
% One of the reasons why SBFL does not work well on \textsc{ZipMe} is that its  test suites quality. 
Particularly, the average test coverage of a variant in this system is only about 43\% (Table~\ref{table:subject_systems}).
Although the baseline performance with Op2 in the system \textsc{ZipMe} is worse than in the other systems, it is still better than other SBFL metrics in \textsc{ZipMe}.

\begin{figure}
    \centering
    \includegraphics[width=0.7\linewidth]{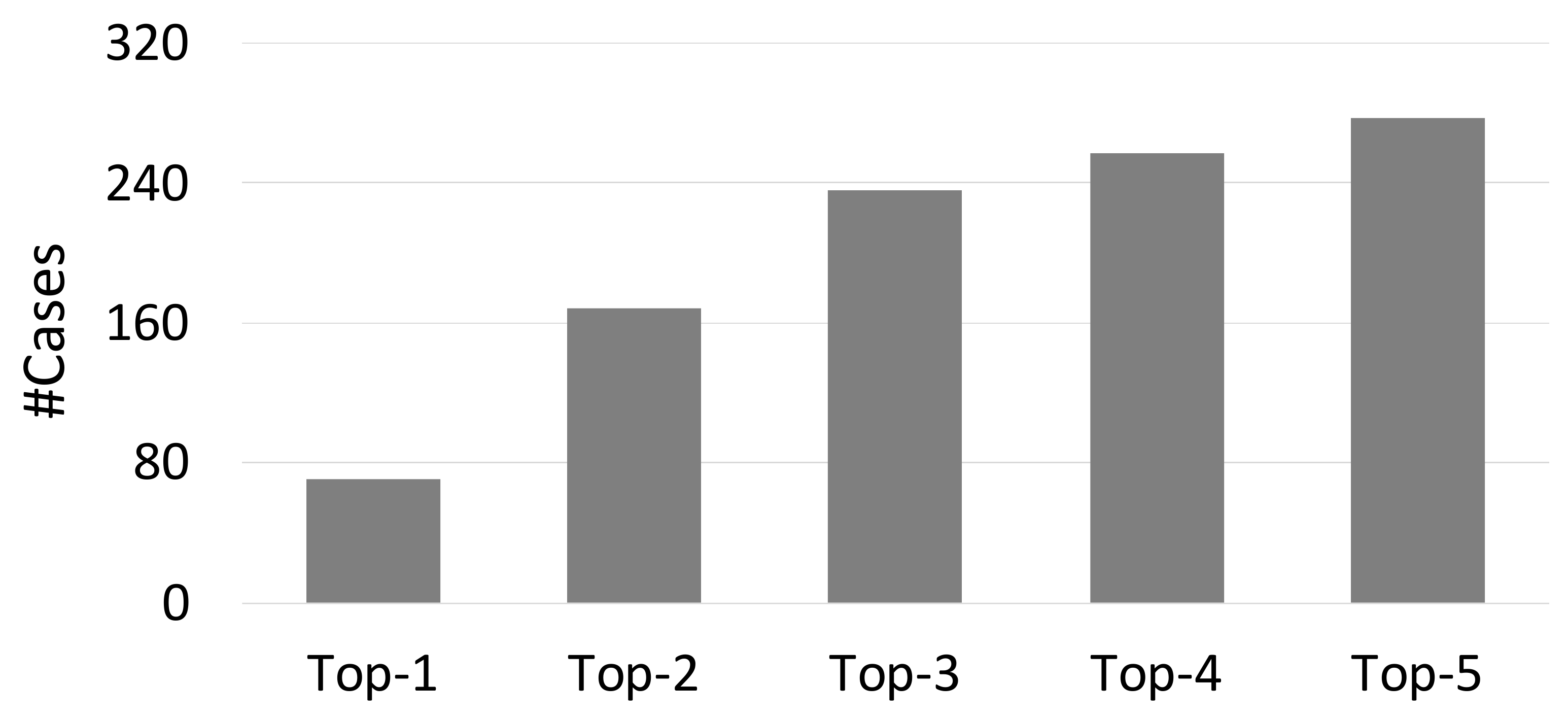}
    \caption{Top-$N$, $N \in [1,5]$ of SBFL with Op2 in Single-bug Setting}
    \label{fig:top_n_single_bugs}
\end{figure}

\begin{figure}
    \centering
    \includegraphics[width=0.7\linewidth]{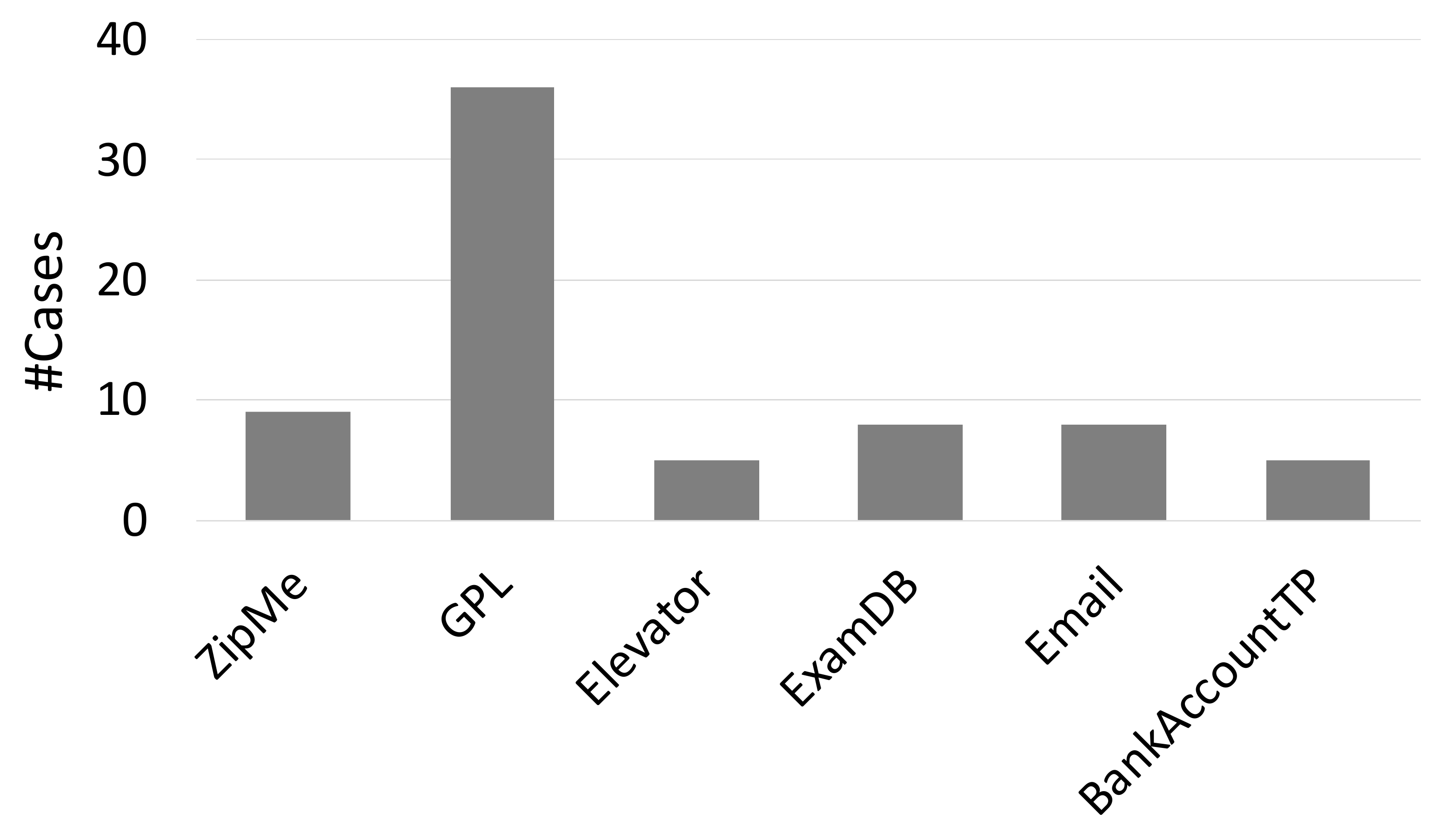}
    \caption{Top-1 of SBFL with Op2 in Single-bug Setting}
    \label{fig:top_1_single_bugs}
\end{figure}

The baseline performance with Op2 in \textit{Top-$N$} is illustrated in Figure~\ref{fig:top_n_single_bugs}. Overall, 71 bugs (about 21\% of the bugs) are correctly ranked at \textit{Top-$1$}. 
In addition, the number of the detected bugs gradually increases when more statements in the ranked lists are examined, about 82\% of the bugs are ranked at \textit{Top-$5$} accuracy.
Interestingly, there are more bugs correctly ranked at the \textit{Top-$1$} positions in \textsc{GPL} than in the other systems (Figure~\ref{fig:top_1_single_bugs}). This is because, for \textsc{GPL}, there is a large number of the sampled products, which provides more information to locate the bugs. Indeed, among the 71 bugs correctly ranked first, there are 36 cases are buggy versions of \textsc{GPL}.

\subsubsection{Performance in localizing multiple bugs}

% To evaluate the baseline performance in localizing multiple bugs, 
We conducted an experiment on 1,232 cases where each buggy version contains $n$ bugs, $n > 1$. Table~\ref{table:baseline_SBFL_multiple_bugs} shows that Ochiai is the most effective metric for localizing multiple bugs in our dataset. On average, with Ochiai, developers only need to investigate about 4 statements in the ranked lists to find the first buggy statement of a faulty system. Meanwhile, this figure with other SBFL metrics is much worse, e.g., with Op2, which is about 13 statements.

\begin{figure}
    \centering
    \includegraphics[width=0.73\linewidth]{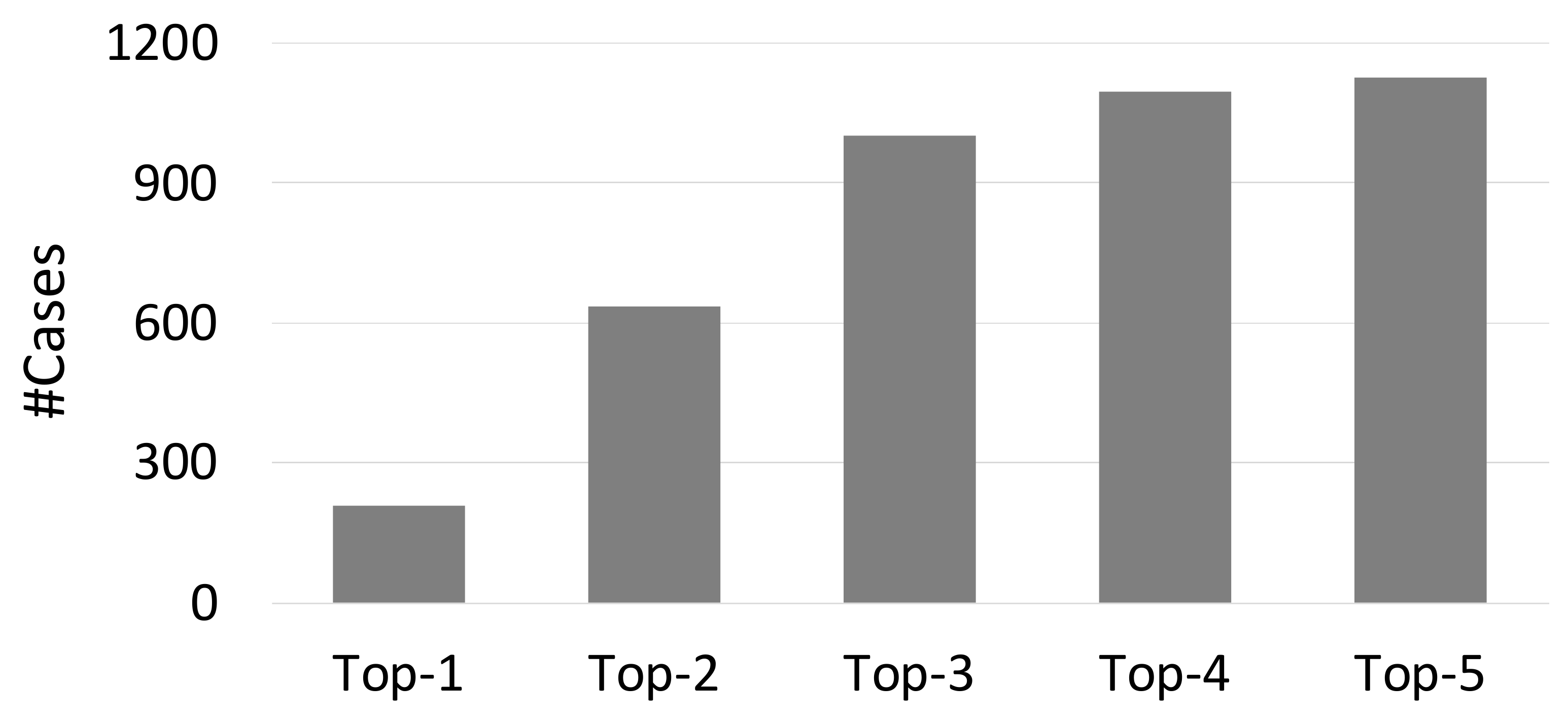}
    \caption{Top-$N$, $N \in [1,5]$ of SBFL with Ochiai in Multiple-bug Setting}
    \label{fig:top_n_multiple_bugs}
\end{figure}

\begin{figure}
    \centering
    \includegraphics[width=0.7\linewidth]{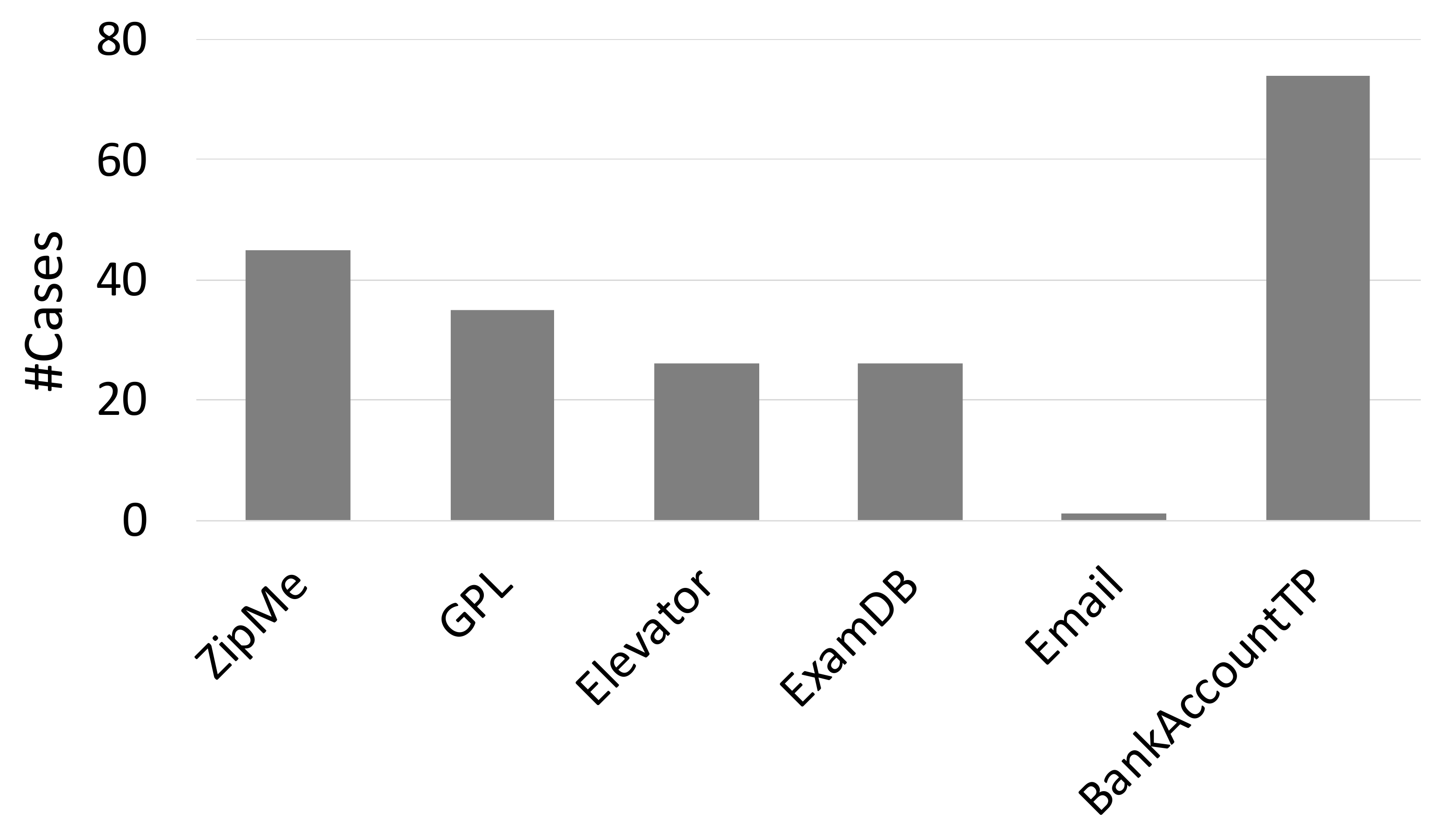}
    \caption{Top-1 of SBFL with Ochiai in Multiple-bug Setting}
    \label{fig:top_1_multiple_bugs}
\end{figure}

In addition, the performance with Ochiai on \textit{Top-$N$} in the multiple-bug setting is shown in Figure~\ref{fig:top_n_multiple_bugs} and Figure~\ref{fig:top_1_multiple_bugs}. Overall, about 17\% and 92\% of the cases have at least one bug ranked at the \textit{Top-$1$} and \textit{Top-$5$} positions, respectively. Among the subject systems, \textsc{Email} is the system in which SBFL with Ochiai achieved the lowest performance in \textit{Top-$1$} (Figure~\ref{fig:top_1_multiple_bugs}). Especially, by examining the first statement, there is only one case that found the bug.

\begin{figure}
    \centering
    \includegraphics[width=0.7\linewidth]{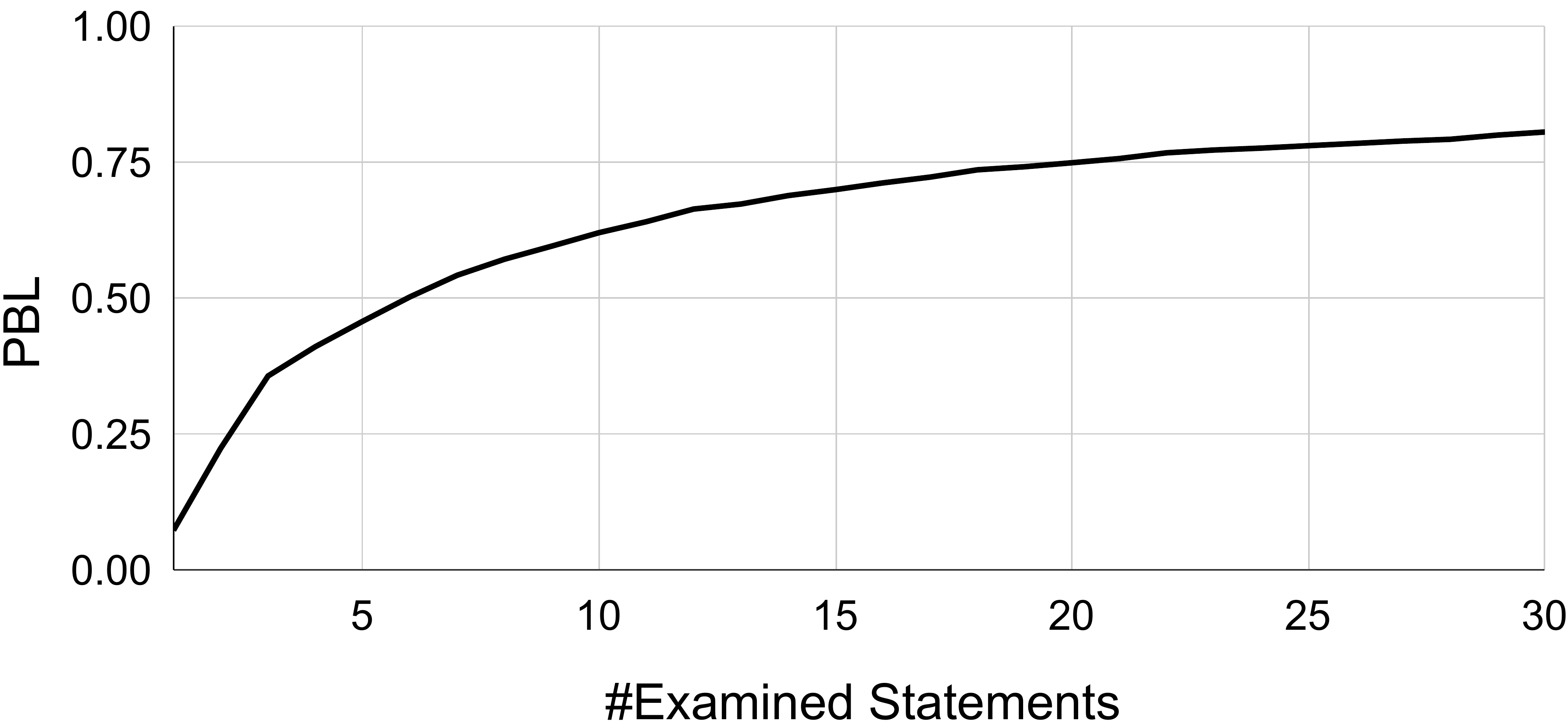}
    \caption{PBL of SBFL with Ochiai in Multiple-bug Setting}
    \label{fig:pbl}
\end{figure}

Furthermore, the average proportion of bugs that are localized in each case by SBFL with Ochiai is shown in Figure~\ref{fig:pbl}. On average, only 7\% of the bugs can be found after examining the first statement. Moreover, by investigating the first 5 statements in the ranked lists, developers can find 46\% of the bugs in a system. Additionally, in order to find about 80\% of the bugs, they need to examine up to 30 statements in the ranked lists.

\begin{table*}
\caption{SBFL Performance in Multiple-bug Setting}
\label{table:baseline_SBFL_multiple_bugs}
  \begin{tabular}{l|rr|rr|rr|rr|rr|rr}
  \toprule
\multirow{2}{*}{\textbf{Metric}} & \multicolumn{2}{c|}{\textsc{\textbf{ZipMe}}} & \multicolumn{2}{c|}{\textsc{\textbf{GPL}}} & \multicolumn{2}{c|}{\textsc{\textbf{Elevator}}} & \multicolumn{2}{c|}{\textsc{\textbf{ExamDB}}} & \multicolumn{2}{c|}{\textsc{\textbf{Email}}} & \multicolumn{2}{c}{\textsc{\textbf{BankAccountTP}}} \\
 & Rank & EXAM & Rank & EXAM & Rank & EXAM & Rank & EXAM & Rank & EXAM & Rank & EXAM \\
 \midrule
    Taratula & 9.04 & 0.39 & 4.08 & 0.42 & 7.98 & 1.78 & 8.21 & 3.27 & 3.53 & 1.43 & 4.92 & 6.36\\
    \textbf{Ochiai} & \textbf{5.07} & \textbf{0.22} & \textbf{2.76} & \textbf{0.29} & \textbf{4.84} & \textbf{1.08} & \textbf{2.75} & \textbf{1.09} & \textbf{3.10} & \textbf{1.26} & \textbf{2.58} & \textbf{3.34} \\
    Op2 & 19.91 & 0.85 & 4.60 & 0.48 & 15.93 & 3.56 & 11.48 & 4.57 & 11.73 & 4.75 & 4.24 & 5.48\\
    Barinel & 9.04 & 0.39 & 4.11 & 0.43 & 7.98 & 1.78 & 8.21 & 3.27 & 3.53 & 1.43 & 4.93 & 6.36\\
    Dstar & 7.58 & 0.32 & 2.79 & 0.29 & 4.78 & 1.07 & 2.74 & 1.09 & 2.99 & 1.21 & 2.65 & 3.44 \\
  \bottomrule
    \multicolumn{13}{l}{\footnotesize \textsc{\textbf{Elevator}}, \textsc{\textbf{Email}} are the abbreviations for \textsc{\textbf{Elevator-FH-JML}} and \textsc{\textbf{Email-FH-JML}} systems, respectively.}
  \end{tabular}
\end{table*}

\section{Summary}
We present a variability fault localization benchmark with a dataset of 1,570 buggy versions of 6 widely-used SPL systems. In our dataset, there are 338 cases of single-bug and 1,232 cases of multiple-bug. The variability bugs in the benchmark are systematically generated by diverse mutation operators, numerous code elements, and different numbers of involving features. We also provide several standard metrics which are broadly applied in evaluating FL techniques. Furthermore, a naive solution adapting SBFL for variability fault localization is evaluated to present the baseline results. 
% Several datasets of variability bugs in SPL systems have been proposed. However, these datasets are designed in a different setting and do not contain the corresponding tests for each bug. 
We hope that our benchmark can be a common point of comparison for variability fault localization techniques and encourage the researchers to propose better solutions for the challenge case.
\begin{acks}
In this work, \textit{Kien-Tuan Ngo} was funded by Vingroup Joint Stock Company and supported by the Domestic Master/ PhD Scholarship Programme of Vingroup Innovation Foundation (VINIF), Vingroup Big Data Institute (VINBIGDATA), code VINIF.2020.ThS.04.

This work has been supported by Vietnam National University, Hanoi (VNU), under Project No. QG.18.61.
\end{acks}

\bibliographystyle{ACM-Reference-Format}
\bibliography{references}

\end{document}